\documentclass[twocolumn,showpacs,preprintnumbers,amsmath,amssymb,floatfix,prl]{revtex4}
\usepackage{graphicx}% Include figure files
\usepackage{dcolumn}% Align table columns on decimal point
\usepackage{bm}% bold math

\newcommand{\beq}{\begin{equation}}
\newcommand{\eeq}{\end{equation}}
\newcommand{\bey}{\begin{eqnarray}}
\newcommand{\eey}{\end{eqnarray}}

\begin{document}

\title{About influence of gravity on heat conductivity process of the Planets}

\author{S.O. Gladkov}
\email{sglad@newmail.ru}
\affiliation{Moscow Aviation Institute, National Research
University, Volokolamskoe Shosse 4, Moscow, Russia}

\author{Anil Yadav}
\email{abanilyadav@yahoo.co.in}
\affiliation{Department of Physics, Galgotias College of
Engineering and Technology, Greater Noida  201306, India}

\author{Saibal Ray}
\email{saibal@iucaa.ernet.in}
\affiliation{Department of Physics, Government College of
Engineering \& Ceramic Technology, Kolkata 700010, West Bengal,
India}

\author{F. Rahaman}
\email{rahaman@iucaa.ernet.in}
\affiliation{Department of Mathematics, Jadavpur University,
Kolkata 700032, West Bengal, India}

\date{today}

\begin{abstract}
In the present study it is shown that the interaction of a
quasi-static gravitational wave through density fluctuations gives
rise to a heat conductivity coefficient and hence temperature.
This fact is a very important characteristics to establish a heat
equilibrium process of such massive body as the Earth and other
Planets. To carry out this exercise general mechanism has been
provided, which makes a bridge between classical physics and
quantum theory, and specific dependence of heat conductivity
coefficient in wide region is also calculated.
 \end{abstract}

%\keywords{Gravity; Coefficient of heat conductivity; Temperature; Massive body}


 \maketitle

%\section{Introduction}
Problems related to study of heat conduction process in the various
bodies an enormous number of attempts are reported in the literature
(see, for example, the original paper \cite{Akhiezer1960,Zubarev1960,Gladkov1992}
and the monograph \cite{Carslaw1960,Misnar1968}). In this case a
single distinguishing feature among all of them is a model
representation of the structure of the body. If it is a crystalline
insulator, then Debye's phonon model does `work'. On the other hand,
for a complex heterogeneous structure, which are in particular
porous body, then the Debye's model as well as the gas approximation is applicable.
This allows us not only to do analytical description correctly and proper implementation
to several experiments \cite{Gladkov2002}), but also it predicts number of additional
effects \cite{Gladkov1990} (as reflected in \cite{Gladkov2003} and 
later on confirmed by experiments \cite{Abraytis1999,Grishin2002}).

In this paper we investigate an influence on the process
of heat transfer mechanisms associated with the interaction of
gravitation via the density fluctuations. As we shall see later,
it is extremely important for any massive body, such as Earth, to account for
this type of effect. As an object of present research, we choose
the simplest stuff and confine ourselves within the dielectric composition
of the matter. To determine the contribution of gravity to the coefficient of thermal
conductivity, one should note that if there are multiple
mechanisms with different physical nature, then the thermal
conductivity becomes an additive function and is determined by the
sum of contributions of relevant mechanisms. For dielectrics
such contribution will consist of only two components and can be represented
as
\begin{eqnarray}
\kappa = \kappa_p+\kappa_{\text{g}},
\end{eqnarray}
where $\kappa_p$ is the coefficient of thermal conductivity due
to density fluctuations (in the Debye's model of phonons) and an
additional contribution  $\kappa_{\text{g}}$ is the coefficient of
thermal conductivity due to the gravity.

%\section{The model and thermal conductivity of massive body}

If the temperature behavior is understood, then its dependence can be
determined by the formula
\begin{equation}
\kappa_{p}= \left\{\begin{array}{rcl}
{AT^3} & \mbox{at}  & T \ll \overline{\theta}_D \\
B & & \\
\overline{T} & \mbox{at} & T \gg \overline{\theta}_D
\end{array}\right.
\end{equation}
Here $A$ and $B$ are well-defined coefficients (for more details one
can look at the Ref. \cite{Gladkov2004} or the classical monographs
\cite{Lifshitz1971,Lifshitz1979}), and $\overline{\theta}_D$
is the average energy of thermal vibrations (for
crystalline body this is simply the Debye temperature).

To calculate the coefficient of gas we use the simplest
approximation and represent it in the form
\begin{equation}
\kappa_{\text{g}}=\frac{1}{3}C_{\text{g}}c^2\tau_{\text{g}-p},
\end{equation}
where $c$ is the speed of light, $C_{\text{g}}$ is the heat capacity of gravity waves per unit
volume of the body and $\tau_{\text{g}-p}$ is the relaxation time
of the transfer of energy between the gravitational wave and phonons.

Most effective mechanism of phonon interaction with the gravitational wave
is the process of disintegration of the gravitational wave into two phonons.
This act of interaction is characterized by the following two conservation laws:
       
        \noindent  
       (i) The law of conservation of energy
       \begin{equation}
        \hbar \omega_{\text{g}}(k)=\hbar [\omega_{p}(q_1) + \omega_{p}(q_2)],
     \end{equation}
      ii) The law of conservation of momentum
    \begin{equation}
     \overrightarrow{k} = \overrightarrow{q}_1 + \overrightarrow{q}_2,
    \end{equation}
    where $\hbar$ is the Planck constant, $\overrightarrow{k}$ and $\overrightarrow{q}_{1,2}$ are the wave vectors of
   gravitational waves and phonon wave respectively. The laws of
   dispersion (frequency dependence of the wave vector) are defined as
   $\omega_{\text{g}}(k) = ck$ and $\omega_{p}(q) = c_sq$, where $c_s$ is the
   average speed of sound for the Earth.

   The simplest algebraic analysis of conservation laws (4) - (5) leads
   to the conclusion that the wave vector of the virtual phonon
   must be enclosed in a narrow range, such that
   \begin{equation}
   \frac{k}{2} \left(\frac{c}{c_s} -1\right) \leq q_1 \leq \frac{k}{2} \left(\frac{c}{c_s} + 1 \right).
   \end{equation}

   Since the speed of light is much greater than the speed of sound,
   then (6) reduces to $q_1 \approx q_2 \approx kc/2c_s$.
   This is the energy of the gravitational wave which is approximately equally distributed
   between the two virtual phonons.

   To write down properly the expression for amplitude of the static gravitational field,
    we must remember that the gravitational potential for distances $r \geq R$, where $R$
     is radius of the body, can be used in the following form \cite{Landau1973}:
     \begin{equation}
     \varphi=\frac{r_{\text{g}}c^2}{2r},
     \end{equation}
     where $r_{\text{g}}$ is the gravitational radius of the planet under consideration, 
     as defined  by $r_{\text{g}}=2GM/c^2$, here $G$ and $M$ are the universal constant of 
     gravitation and mass of planet respectively.

     The transition from classical expression (7) to quantum representation is now a
     fundamental requirement which can be obtained through the Navigation Rules \cite{Gladkov1999}.
     Then the second quantization formalism in the quasi-static approximation can be written as
     \begin{eqnarray}
     \hat{\varphi}(\overrightarrow{r},t)=i\sum_k \sqrt\frac{2\pi G
     \hbar c}{V_{\text{g}} k} \times \nonumber \\
     \left[{\hat{c}_k^+} e^{-i(\overrightarrow{k}.\overrightarrow{r} - \omega_{\text{g}} t)} +
     \hat{c_k} e^{i(\overrightarrow{k}.\overrightarrow{r} - \omega_{\text{g}}
     t)}\right],
     \end{eqnarray}
     where $\hat{c}_k (\hat{c}_k^+)$ is the annihilation operator
     (creation) of gravitational waves and $V_{\text{g}}=4\pi r_{\text{g}}^3/3$ is the quantization volume of
      corresponding gravitational field.

     To get the Hamiltonian of the interaction between density
     fluctuations and gravitational waves, we shall now employ some
     basic postulates of differential geometry. Indeed, in a curved space
     the volume integral can be represented in the following invariant form:
     \begin{equation}
     J=\int_V(...)\sqrt{-g}dV,
     \end{equation}
     where $g$ is determinant of the metric tensor. Here `minus' sign characterizes the
     pseudo-Euclidean space-time. In almost static gravitational field
     the time component of the metric tensor is given by \citep{Landau1973}
     \begin{equation}
     g_{00} \approx 1- \frac{2\varphi}{c^2}.
     \end{equation}
     The above indices run upto four values and the index corresponds to the time coordinate.
     Actually here the situation demands that $\sqrt{-g} = \sqrt{-(g_0+\delta{g})} 
     \approx \sqrt{-g_0}\left(1+ \frac{\delta{g}}{2g_0}\right)$,
      where $\delta{g} \ll g_0$, as in a flat pseudo-Euclidean spacetime we get here the similar result.

      This means that the transition to the non-relativistic approximation of interaction
      with an almost static gravitational field must be described by an expression in
      operator form. Thus the desired interaction of gravitational waves with the
      density fluctuations, whose role in the formal language of phonons is played
      by the strain tensor, is given by
      \begin{equation}
       \hat{H}_{int}=\frac{\overline{\theta}_D}{Vc^2} \int_V \hat{u}_{\alpha\beta}^2\hat{\varphi}dV,
       \end{equation}
       where $a$ is the average distance between nodes of localized
       oscillating structures on the substance and is related to $\overline{\theta}_D $ as follows:
       $\overline{\theta}_D = \hbar\overline{c}_s/\overline{a}$.

       The strain tensor $u_{\alpha\beta}$ associated with the displacement vector points of
       the continuum ratio $u_{\alpha}$ can be written by
       \begin{equation}
        u_{\alpha\beta}=\frac{1}{2}\left(\frac{\partial{u_{\alpha}}}{\partial{x_{\beta}}} 
      + \frac{\partial{u_{\beta}}}{\partial{x_{\alpha}}}\right).
    \end{equation}

 Again, secondary-quantized expression for the displacement vector has the following form
\begin{eqnarray}
\hat{u}_{\alpha}=\sum_{q,e} e_{\alpha} \left(\frac{\hbar}{2\rho
V\omega_{p}(q)}\right)^{\frac{1}{2}} \times \nonumber \\
\left[{\hat{b}}_q^+
e^{-i(\overrightarrow{q}.\overrightarrow{r} - \omega_{p} t)} +
\hat{b}_q e^{i(\overrightarrow{q}.\overrightarrow{r} - \omega_{p}
t)}\right],
\end{eqnarray}
where $\rho$ is the density structure, $V$ is its volume, whereas
$\hat{b}_q^+(\hat{b}_k)$ is the creation (annihilation) of a
phonon with wave vector $\overrightarrow{q}$ and
$\overrightarrow{e}_{\alpha}$ is the polarization vector of the
sound wave.

Now if we substitute above Eq. (13) into Eq. (12), and it in turn to the
Hamiltonian (11), then considering Eq. (9) we find the expected expression
for interaction of energy density fluctuations with the gravitational field as follows:
 \begin{eqnarray}
 \hat{H}_{int}=\frac{\pi \overline{\theta}_D \hbar}{2V \rho c_s c^2}\sqrt{\frac{2\pi G c \hbar}{V_{\text{g}}}} \times \nonumber \\
 \sum_{k,q_1,q_2}\frac{(\overrightarrow{e}_1.\overrightarrow{q}_1)(\overrightarrow{e}_2.\overrightarrow{q}_2)}{\sqrt{kq_1q_2}}
\left[\hat{b}_{q_1}^+ - \hat{b}_{-q_1}\right] \left[\hat{b}_{q_2}^+ - \hat{b}_{-q_2}\right] \times \nonumber \\
\left[\hat{c}_k^+ + \hat{c}_{-k}\right]  \Delta[\overrightarrow{k} + \overrightarrow{q_1} + \overrightarrow{q_2}],
 \end{eqnarray}
where $\Delta$ is function of $x$ such that $\Delta(x)=1$ if $x=0$ and $\Delta(x)=0$ for all $x \neq 0$.

Knowing the interaction (14) it is now easy to find out an expression for relaxation
time $\tau_{\text{g}-p}$  of the gravitational wave appearing in the formula (3). In
fact, if we use the recipe described in detail in Ref. \citep{Gladkov1999}, after
some simple calculations taken into account, the conservation laws (4)
and (5) through (6) will provide the desired decay time of the
gravitational wave into two phonons:
\begin{equation}
\frac{1}{\tau_{\text{g}-p}(k)} \approx \frac{\pi^2}{32} \left(\frac{V}{V_g}\right) \frac{G\overline{\theta}_D^2}{\hbar c_s^5}
\left(\frac{\hbar}{\rho c_s \overline{a}^4}\right)^2 (\overline{a} k)^2kc(1+2\overline{N}_k),
 \end{equation}
 where $\overline{N}_k=\frac{1}{e^{\hbar c_s k/T} - 1}$ is the equilibrium phonon distribution function
 of Bose-Einstein statistics in standard form. The Boltzmann constant here after to be
 equal to unity. Note that in (15) there was a big factor $V/V_{\text{g}}$ due to transitions from
 summation to integration over the region of phase space phonon. It can be shown in the form
 $$\sum_q f(q)=V\int_V f(q) \frac{d^3q}{(2\pi)^3}.$$

 For averaging formula (15) on the distribution of Planck statistics,
 which are subject to gravitational waves, we use the following simple rule:
 $$\overline{(...)}=\frac{\int_0^{\infty}k^2 \overline{f}_k(...)dk}{\int_0^{\infty}k^2 \overline{f}_k dk},$$
  where $ \overline{f}_k = \left(e^{\hbar c k/T} - 1\right)^{-1}$.

    The above rule leads us to the following expression
    for the average relaxation time of the gravitational wave
    \begin{equation}
    \frac{1}{\overline{\tau}_{\text{g}-p}} \approx \frac{\pi^2}{32} \left(\frac{V}{V_{\text{g}}}\right) \frac{G\overline{\theta}_D^2c}{\hbar c_s^5\overline{a}}
    \left(\frac{\hbar}{\rho c_s \overline{a}^4}\right)^2 \left(\frac{\overline{a}T}{\hbar c}\right)^3 (1+2\overline{N}(u)),
     \end{equation}
     where $u=c_s/c$ and for the constraint $u\ll1$, we get $\overline{N}(u) \approx 1/u =c/c_s$.

     Therefore, under the above approximation for a spherical body one can assume the equation (16) in its modified form
     \begin{equation}
     \frac{1}{\overline{\tau}_{\text{g}-p}} \approx \frac{\pi^2}{16} \frac{G\overline{\theta}_D^2c^2}{\hbar c_s^6\overline{a}}
     \left(\frac{\hbar}{\rho c_s \overline{a}^4}\right)^2 \left(\frac{\overline{a}T}{\hbar c}\right)^3 \left(\frac{R}{r_{\text{g}}}\right)^3.
      \end{equation}

   Since the heat capacity of the particles with a linear range of
the wave vector (as well as in the Debye model) is proportional
to $T^3$, then according to Eqs. (17) and (3) we have
\begin{equation}
\kappa_{\text{g}}=\frac{1}{3}C_{\text{g}}c^2\tau_{\text{g}-p} \sim \left(\frac{T}{\hbar c}\right)^3 c^2 \overline{\tau}_{\text{g}-p}=constant= D,
\end{equation}
where the new coefficient $D$ is defined as
\begin{equation}
D \approx \left(\frac{c_s}{c}\right)^3 \left(\frac{\rho c_s \overline{a}^4}{\hbar}\right)^2
\frac{\hbar c_s^6}{G\overline{a}^2 \overline{\theta}_D^2} \left(\frac{r_{\text{g}}}{R}\right)^3.
 \end{equation}

Therefore, by summing Eqs. (2) and (18), we find the expression for thermal
conductivity for massive structure as follows
 \begin{equation}
 \kappa = \left\{\begin{array}{rcll}
 {AT^3} & \mbox{if} & T \ll \overline{\theta}_D & \\
 B & & & +D \\
 \overline{T} & \mbox{if} & T \gg \overline{\theta}_D &
 \end{array}\right.
 \end{equation}

At this point it would be very appropriate to give an estimate of average relaxation time
of the quasi-static gravitational wave. Hence, according to Eq. (17), we obtain the order of magnitude for relaxation time as
\begin{equation}
\overline{\tau}_{\text{g}-p} \approx 10^{-8}c. 
\end{equation} 
For the above estimation we have used the following numerical values of the parameters:
$\hbar=10^{-27} erg-s,~G=6.67\times10^{-8}~CGS,~\overline{\theta}_D=100K,
~c=3\times10^{10}~cm-s^{-1},~c_s=10^5~cm-s^{-1},~\rho=5.5~g-cm^{-3},
~\overline{a}=10^{-8}~cm,~R=6.4\times10^8~cm,~r_{\text{g}}=10~cm,~T=300~K=3\times10^{-14}~erg. $

Note that above value of relaxation time represents a very fast thermalization process
of gravitational wave and points to the need for a massive body.

%\section{Concluding remarks}

To summarize, in the present paper we have studied influence of gravity in the thermal
conductivity process of massive bodies. Actually the interaction of a quasi-static gravitational
wave via density fluctuations gives rise to a heat
conductivity coefficient $\kappa$ in connection to a temperature $T$.
Therefore, it is important to establish a heat equilibrium process
of such massive body like the Earth and other Planets.

As a major result of the investigation we would like to highlight the following two important points:

 (1) In the description of thermal conductivity of massive bodies it is extremely important
 to consider the interaction of gravitational waves with the remaining substance, however its composition
will be determined by the particular material of the structure.

 (2) The coefficient of thermal conductivity of such bodies due to
gravitational waves increases with temperature which eventually gets
saturated and tends to a constant value such that at high temperature it ceases to
depend on any external influence.

%\section*{Acknowledgments}
SR and FR are thankful to the authority of Inter-University
Centre for Astronomy and Astrophysics, Pune, India for providing
Visiting Associateship under which a part of this work was carried
out.

\end{document}